\documentclass[twocolumn,showpacs,amsmath,amssymb,aps,prl]{revtex4}

\usepackage{graphicx}

\usepackage{amsmath}

\newcommand{\prt}{\partial}
\newcommand{\II}{\mbox{${\mathbb I}$}}

\newcommand{\RR}{\mbox{${\mathbb R}$}}

\def\UU{\mathbb U}
\def\S{\mathbb S}

\newcommand{\rd}{{\rm d}}

\newcommand{\tx}{\widetilde{x}}

\def\ri{{\rm i}}

\def\prt{{\partial}}

\def\e{{\rm e\, }}

\def\S{{\mathbb S}}
\def\ri{{\rm i}}

\def\t0{\tau_0}

\newcommand{\be}{\begin{equation}}
\newcommand{\ee}{\end{equation}}
\newcommand{\bea}{\begin{eqnarray}}
\newcommand{\eea}{\end{eqnarray}}

\begin{document}


\title{Thermoelectric efficiency of critical quantum junctions} 

\author{
Mihail Mintchev$^{1,2}$, Luca Santoni$^2$ and Paul Sorba$^3$
}
 \affiliation{
 ${}^1$ Istituto Nazionale di Fisica Nucleare, Largo Pontecorvo 3, 56127 Pisa, Italy \\
${}^2$ Dipartimento di Fisica dell'Universit\`a di Pisa, 
Largo Pontecorvo 3, 56127 Pisa, Italy\\ 
${}^3$ LAPTh, Laboratoire d'Annecy-le-Vieux de Physique Th\'eorique, 
CNRS, Universit\'e de Savoie,   
BP 110, 74941 Annecy-le-Vieux Cedex, France}

\date{\today}

\begin{abstract}

We derive the efficiency at maximal power of a scale-invariant 
(critical) quantum junction in exact form. Both Fermi and Bose statistics are considered. We show 
that time-reversal invariance is spontaneously broken. For fermions we implement a new 
mechanism for efficiency enhancement above the Curzon-Ahlborn bound, 
based on a shift of the particle energy in each heat reservoir, proportional to its temperature. 
In this setting fermionic junctions can even reach at maximal power the Carnot efficiency. 
Bosonic junctions at maximal power turn out to be less efficient then fermionic ones.

\end{abstract}
\maketitle

There is recently much interest in the study of thermoelectric phenomena in nanoscale 
devices and in particular, in nanoscale engines. The efficiency of such engines is 
a fascinating physical problem. As is well known, one relevant 
parameter for studying this problem is the efficiency $\eta(P_{\rm max})$ at maximal power 
$P_{\rm max}$. In the context of classical linear endoreversible thermodynamics (irreversible 
heat transfer) it has been shown \cite{ca, vb} that 
\begin{equation}
\eta(P_{\rm max}) \leq 1-\sqrt{\frac{T_2}{T_1}} \equiv \eta_{\rm CA}\, ,
\label{CA}
\end{equation} 
where $T_1>T_2$ are the temperatures of the two reservoirs, needed for running the engine. 
The inequality (\ref{CA}) is known as Curzon-Ahlborn (CA) bound. In a series of recent 
papers \cite{bsc}-\cite{bs} it has been 
proposed that at the quantum level $\eta(P_{\rm max})$ might be enhanced in principle above 
$\eta_{\rm CA}$ by means of an {\it explicit} breaking of time-reversal symmetry. For this purpose, 
the authors of \cite{bsc}-\cite{bbc} considered in the linear response regime a three-terminal setup with 
one probe terminal and a magnetic field, which breaks down time-reversal. 
A generalization of this idea to multi-terminal systems has also been studied \cite{bs}. 

In the present paper we investigate the efficiency of quantum Schr\"odinger junctions 
with both Fermi and Bose statistics. We demonstrate that when the interaction, driving the system away 
from equilibrium is {\it scale invariant} (critical), one can go 
beyond the linear response approximation and derive $\eta(P_{\rm max})$ in exact form. 
Time reversal invariance is {\it spontaneously} broken, which provides in the quantum world 
an attractive alternative to the explicit breaking in \cite{bsc}-\cite{bbc}. For fermions 
we propose and investigate a new mechanism for efficiency 
enhancement above $\eta_{\rm CA}$, based on a shift of the energy in the heat reservoirs 
proportional to their temperature. With an appropriate shift, fermionic junctions can reach at 
maximal power even the Carnot efficiency $\eta_{\rm C}$. Analogous behavior has been 
observed \cite{s} in the stochastic model of an isothermal engine. 
At maximal power the bosonic junctions are less efficient and do not attain $\eta_{\rm C}/2$. 

\begin{figure}[ht]
\begin{center}
\begin{picture}(80,40)(220,300)
\includegraphics[scale=0.85]{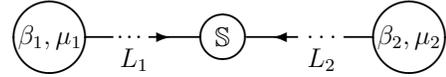}
\end{picture} 
\caption{(Color online) Schematic picture of the junction with two heat reservoirs connected via  
two leads $L_1$ and $L_2$ to the scattering matrix $\S$.} 
\label{fig1}
\end{center}
\end{figure}

{\it The system}: The scheme of the junction, considered in this Letter, is shown in Fig. \ref{fig1}. 
The two thermal reservoirs at (inverse) temperature $\beta_i$ and chemical potential $\mu_i$ are 
connected with two semi-infinite leads through a point-like interaction characterized by a unitary 
scattering matrix $\S$. The leads $L_i$ are modeled by two half-lines with local 
coordinates $\{(x,i)\, :\, x<0,\, i=1,2\}$, $\S$ being localized at $x=0$. 
The system is away from equilibrium provided 
that $\S$ has a non-vanishing transmission amplitudes and $\beta_1$ and/or $\mu_1$ 
differ from $\beta_2$ and/or $\mu_2$. The dynamics is fixed by the Schr\"odinger equation 
\begin{equation}
\left (\ri \prt_t +\frac{1}{2m} \prt_x^2 - \frac{a}{\beta_i} \right )\psi (t,x,i) = 0\, , 
\label{eqm}
\end{equation} 
where $m$ is the mass and $a$ is a dimensionless real parameter. We show in what follows 
that the term $a/\beta_i$ (a temperature dependent potential), 
generating a shift in the dispersion relations 
\begin{equation}
\omega_i(k) = \frac{k^2}{2m} +\frac{a}{\beta_i}  
\label{dr}
\end{equation} 
of the particles in the two heat baths, affects $\eta(P_{\rm max})$. 

We adopt a field theory formulation. Accordingly, the Schr\"odinger field $\psi(t,x,i)$ with (Fermi) Bose  
statistics satisfies the standard equal-time canonical (anti)commutation relations. 
The interaction at the point $x=0$ is fully codified in the boundary condition 
\begin{equation} 
\lim_{x\to 0^-}\sum_{j=1}^2 \left [\varrho (\II-\UU)_{ij} +\ri (\II+\UU)_{ij}\prt_x \right ] \psi (t,x,j) = 0\, , 
\label{bc} 
\end{equation} 
where $\UU$ is an arbitrary $2\times 2$ unitary matrix and $\varrho \in \RR$ is a 
parameter with dimension of mass. This is \cite{ks-00,h-00} the most general 
boundary condition, implying the self-adjointness of the operator $-\prt_x^2$ and thus of the 
Hamiltonian of the system. The scattering matrix, associated with the point-like interaction 
generated by (\ref{bc}), is \cite{ks-00,h-00}
\begin{equation} 
\S (k) = -\frac{[\varrho (\II - \UU) - k(\II+\UU )]}{[\varrho (\II - \UU) + k(\II+\UU )]}\, ,   
\label{S}
\end{equation} 
which is unitary and satisfies $\S^*(k)=\S(-k)$ (Hermitian analyticity) \cite{f2} and 
$\S(\varrho )=\UU$. Summarizing, the scattering matrix (\ref{S}) describes all possible point-like 
interactions, which generate a unitary time evolution of $\psi$. Junctions with more then two terminals 
can be treated \cite{Bellazzini:2006jb} along the same lines. 

{\it The non-equilibrium steady state $\Omega_{\beta,\mu}$}: Following the pioneering 
work of Landauer \cite{land} and B\"uttiker \cite{butt}, non-equilibrium systems of the 
type shown in Fig. \ref{fig1} have been extensively investigated (see \cite{data} and references therein). 
We use here an algebraic construction \cite{Mintchev:2011mx} of the 
Landauer-B\"uttiker (LB) steady state $\Omega_{\beta,\mu}$ for 
the problem (\ref{eqm}-\ref{S}), allowing to establish explicitly the spontaneous 
breakdown of time-reversal symmetry. Referring for the details to \cite{Mintchev:2011mx}, 
we report only the two-point {\it non-equilibrium} correlation function, needed in what follows. 
Denoting by $\langle \cdots \rangle_{\beta,\mu}$ the expectation value in the state 
$\Omega_{\beta,\mu}$, one has 
\begin{eqnarray}
\langle \psi^*(t_1,x_1,i) \psi(t_2,x_2,j)\rangle_{\beta, \mu} = 
\int_0^{\infty} \frac{\rd k}{2\pi} \e^{\ri \omega_i(k) t_1 -\ri \omega_j(k) t_2 } 
\nonumber \\
\Bigl [\e^{\ri k x_{12}} \delta_{ji} d^\pm_i(k) + \e^{-\ri k x_{12}} \sum_{l=1}^2 \S_{jl}(k) d^\pm_l(k) \S^\ast_{li}(k)  
\qquad \quad 
\nonumber \\
\e^{\ri k \tx_{12}} \S_{ji}(k) d^\pm_i(k)  +  \e^{-\ri k \tx_{12}}  d^\pm_j(k) \S^\ast_{ji}(k) \Bigr ]\, , 
\qquad \qquad 
\label{corr}
\end{eqnarray}
where $x_{12}=x_1-x_2$,  $\tx_{12} =x_1+x_2$ and 
\begin{equation}
d_i^\pm (k) = \frac{1}{\e^{\beta_i[\omega_i(k)-\mu_i]}\pm 1} 
\label{fd}
\end{equation}
is the Fermi/Bose distribution in the $i$-th reservoir. The correlation function (\ref{corr}) is essentially 
the only input for deriving the efficiency $\eta(P_{\rm max})$ below. 

{\it Time-reversal}: It is natural to consider the time reversal symmetry 
as a quantum counterpart of classical reversibility, thus interpreting its breakdown  
as {\it quantum irreversibility}.
The equation of motion (\ref{eqm}) is invariant under the conventional time-reversal operation
\begin{equation}
T \psi (t,x,i) T^{-1} = -\chi_T \psi (-t,x,i)\, , \qquad  |\chi_T|=1\, ,
\label{tr1}
\end{equation}
$T$ being an {\it anti-unitary} operator. The same is true for the boundary condition (\ref{bc}), 
provided that $\UU$ (and therefore $\S(k)$) is symmetric \cite{Bellazzini:2009nk}. 
In spite of the fact that in this case both dynamics and 
boundary condition preserve the time-reversal symmetry, it turns out \cite{Mintchev:2011mx} that the 
LB state $\Omega_{\beta,\mu}$ breaks it down. The simplest way to detect this 
spontaneous breakdown is to use (\ref{corr}) and observe that 
\begin{eqnarray} 
\langle \psi^*(t_1,x_1,i) \psi(t_2,x_2,j)\rangle_{\beta, \mu} \not= \qquad \qquad \qquad 
\nonumber \\
\langle \psi^*(-t_2,x_2,j) \psi(-t_1,x_1,i)\rangle_{\beta, \mu} \, ,
\label{T2}
\end{eqnarray} 
implying $T \Omega_{\beta,\mu} \not= \Omega_{\beta,\mu}$. The above argument shows that 
time-reversal is broken in the LB state $\Omega_{\beta,\mu}$ independently on the presence 
or absence of magnetic field or other explicitly breaking terms. This fact should not be surprising 
because $\Omega_{\beta,\mu}$ is a non-equilibrium state. 

{\it Thermoelectric transport in $\Omega_{\beta,\mu}$}: The particle and energy currents 
are given by 
\begin{equation}
j_x(t,x,i)= \frac{\ri}{2m} \left [ \psi^*(\partial_x\psi ) - (\partial_x\psi^*)\psi \right ]  (t,x,i) \, , 
\label{curr1}
\end{equation}
\begin{eqnarray}
\theta_{xt} (t,x,i) = \frac{1}{4m}  [\left (\partial_t \psi^* \right )\left (\partial_x \psi \right ) 
+ \left (\partial_x \psi^* \right )\left (\partial_t \psi \right ) 
\nonumber \\ - 
\left (\partial_t \partial_x \psi^* \right ) \psi - 
\psi^*\left (\partial_t \partial_x \psi \right ) ](t,x,i) \, . 
\label{en1} 
\end{eqnarray} 
Inserting (\ref{curr1},\ref{en1}) in the correlator (\ref{corr}), one gets in the limit $x_1 \to x_2=x$ 
the Landauer-B\"uttiker expressions 
\begin{eqnarray}
J_i^N \equiv \langle j_x(t,x,i) \rangle_{\beta, \mu} = \qquad \quad 
\nonumber \\
\int_0^\infty \frac{\rd k}{2\pi} \frac{k}{m} \sum_{j=1}^2 \left [\delta_{ij} -  
|\S_{ij}(k)|^2\right ] d^\pm_j(k)\, , 
\label{curr}
\end{eqnarray} 
\begin{eqnarray}
J_i^E \equiv \langle \theta_{xt}(t,x,i) \rangle_{\beta, \mu} = \qquad \quad 
\nonumber \\
\int_0^{\infty} \frac{\rd k}{2\pi}  \frac{k}{m}\, \sum_{j=1}^n \left [\delta_{ij} -  
|\S_{ij}(k)|^2\right ] \omega_j(k) d^\pm_j(k)  \, . 
\label{en}
\end{eqnarray} 
We stress that the expectation values (\ref{curr},\ref{en}) in the state $\Omega_{\beta,\mu}$ 
are exact and satisfy Kirchhoff's rule. No approximations (like linear response theory) have been used. 

{\it Scale invariance}: The $k$-integration in (\ref{curr},\ref{en}) 
with general $\S$-matrix of the form (\ref{S}) cannot 
be performed in closed analytic form. For this reason it is instructive to select among (\ref{S}) 
the {\it scale-invariant} matrices, which incorporate the universal features of the system 
while being simple enough to be analyzed explicitly. These so called {\it critical points} of the set (\ref{S}),  
are fully classified \cite{Calabrese:2011ru}. 
One has two isolated points $\S=\pm \II$ and the family 
\begin{equation}
\S^U = U  \left(\begin{array}{cc} 1 & 0\\0& -1\\ \end{array} \right) U^* \, , \qquad U\in U(2)\, , 
\label{critt}
\end{equation}
which is the orbit of the matrix ${\rm diag}(1,-1)$ under the adjoint 
action of the unitary group $U(2)$. The transport for $\S=\pm \II$ is trivial because in this 
case the leads $L_i$ are actually disconnected. So, we are left with (\ref{critt}) for which the 
$k$-integration in (\ref{curr},\ref{en}) is easily performed. From now on we 
consider the Fermi and Bose statistics separately. 

{\it Exact efficiency-fermions}: With the Fermi distribution and the $\S$-matrix (\ref{critt}) one 
infers from (\ref{curr},\ref{en})
\begin{equation}
J^N_1=\frac{|\S^U_{12}|^2}{2\pi} \left [
\frac{1}{\beta_1} \ln \left (1+ \e^{\beta_1 \mu_1-a} \right ) -
\frac{1}{\beta_2} \ln \left (1+ \e^{\beta_2 \mu_2-a} \right ) \right ]  
\label{curr2} 
\end{equation} 
\begin{eqnarray}
J^E_1 = 
\frac{|\S^U_{12}|^2}{2\pi} \Bigl [
\frac{a}{\beta_1^2} \ln \left (1+ \e^{\beta_1 \mu_1-a} \right ) -
\frac{a}{\beta_2^2} \ln \left (1+ \e^{\beta_2 \mu_2-a} \right ) 
\nonumber \\
-\frac{1}{\beta_1^2}{\rm Li}_2\left (-\e^{\beta_1 \mu_1-a}\right ) +
\frac{1}{\beta_2^2} {\rm Li}_2\left (-\e^{\beta_2 \mu_2-a}\right ) \Bigr ], \qquad 
\label{en2}
\end{eqnarray} 
${\rm Li}_2$ being the dilogarithm function. By Kirchhoff's rule, $J^N_2=-J^N_1$ and  
$J^E_2=-J^E_1$. We stress that at criticality the whole information about 
the interaction, driving the system away from equilibrium, factorizes 
in the transmission probability $|\S_{12}^U|^2$ in front of the expectation 
values of the currents. This remarkable simplification allows us to compute the 
efficiency
\begin{equation}
\eta = \frac{(\mu_2-\mu_1)J^N_1}{J_1^Q} \, , \quad J_1^Q=J_1^E - \mu_1 J_1^N\, , 
\label{eff1}
\end{equation}
exactly, $J_i^Q$ being the heat currents. For this purpose 
we assume $\beta_2>\beta_1$ and 
introduce the variables 
\begin{equation}
\lambda_i = -\beta_i \mu_i\, \qquad r = \beta_1/\beta_2 \in [0,1]\, . 
\label{variables}
\end{equation}
Then, using (\ref{curr2}), the electric power takes the form
\begin{eqnarray}
P(\lambda_1,\lambda_2,r;a) = (\mu_2-\mu_1) J^N_1 = 
\frac{|\S^U_{12}|^2}{2\pi \beta_1^2} (\lambda_1-r \lambda_2)\times \nonumber \\
\left [\ln \left (1+ \e^{-\lambda_1-a} \right )-
r \ln \left (1+ \e^{-\lambda_2-a} \right )\right ] \, . \qquad 
\label{pw1}
\end{eqnarray}
Let us derive now $\eta (P_{\rm max})$. We maximize (\ref{pw1}) by varying 
$\lambda_1$ and $\lambda_2$ for fixed but arbitrary $r$ and $a$. From 
$\prt_{\lambda_1}P = \prt_{\lambda_2} P = 0$ one can deduce that the extrema of 
(\ref{pw1}) are localized at $\lambda_1=\lambda_2\equiv \lambda$, which satisfies 
the $r$-independent equation 
\begin{equation}
\lambda - (1+\e^{\lambda+a}) \ln (1+\e^{-\lambda-a}) =0\, . 
\label{pw2}
\end{equation}
One can also show that for $a\in \RR$ the equation (\ref{pw2}) has a unique solution $\lambda_a$,  
leading to maximal $P$. Inserting this information in (\ref{en1},\ref{eff1},\ref{pw1}), one gets 
\begin{eqnarray}
\eta_f(P_{\rm max}) = \qquad \qquad \qquad \qquad \quad 
\nonumber \\
\frac{(1-r)\lambda_a \ln \left (1+\e^{-\lambda_a-a}\right )}
{(\lambda_a +a+a r)\ln \left (1+\e^{-\lambda_a-a}\right ) -(1+r){\rm Li}_2\left (-\e^{-\lambda_a-a}\right )}\, ,  
\nonumber \\
\label{eff2}
\end{eqnarray}
which represents our main result. Notice that $\eta_f(P_{\rm max})$ vanishes in the isothermal limit $r\to 1$. 

In order to clarify the role of the parameter $a\in \RR$, we investigate the entropy production 
\begin{equation}
\dot{S} \equiv (\beta_2 -\beta_1)J^Q_1 -(\mu_2-\mu_1) \beta_2 J_1^N \, . 
\label{ent1}
\end{equation} 
At maximal power one finds for fermions  
\begin{eqnarray}
\dot{S}(a)= \frac{}{}\frac{|\S^U_{12}|^2(1+r)(1-r)^2}{2\pi r \beta_1}\times \qquad  \nonumber \\
\left [a \ln \left (1+\e^{-\lambda_a-a}\right ) - {\rm Li}_2\left (-\e^{-\lambda_a-a}\right )\right ] \, ,
\label{ent2}
\end{eqnarray}
implying the existence of a point $a_f= -1.1628...$, such that 
$\dot{S}(a) \gtrless 0$ for $a\gtrless a_f$ and $\dot{S}(a_f)=0$. On the other hand, using  
(\ref{en2}) and (\ref{ent2}), one obtains the following relation between 
entropy production and energy flow at maximal power
\begin{equation}
J^E_1(a) = \frac{r}{\beta_1(1-r)}\dot{S}(a)\, .  
\label{new}
\end{equation} 
Combining these results with the orientation of the leads $L_i$ in Fig. \ref{fig1}, 
we conclude that the energy flow is in the direction $1\to 2$ for $a>a_f$ and $2\to 1$ for 
$a<a_f$. Therefore, since $T_1>T_2$, our junction operates as a 
thermoelectric engine for $a>a_f$. It turns out that for $a<a^\prime_f=-3.5890...<a_f$ 
not only the energy flow 
$J_1^E$, but also the heat flow $J_1^Q$ is in the direction $2\to 1$ (for 
any $r\in [0,1]$) and thus our devise works as refrigerator. 

\begin{figure}[ht]
\begin{center}
\begin{picture}(80,100)(55,11)
\includegraphics[scale=0.68]{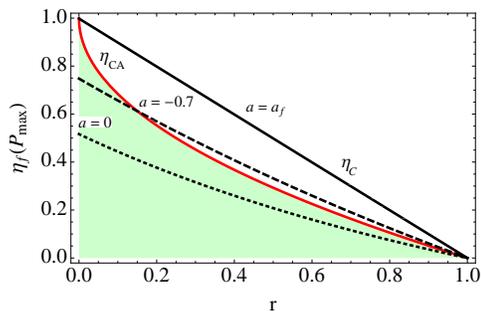}
\end{picture} 
\caption{(Color online) The CA bound (continuous red line) compared to 
$\eta_f(P_{\rm max})$ for $a=0$ (dotted line), $a=-0.7$ (dashed line) and $a=a_f$ (continuous black line).}
\label{fig2}
\end{center}
\end{figure}

Let us study in detail the behavior of the junction as a thermoelectric 
engine. For this purpose one solves numerically 
the equation (\ref{pw2}) for fixed $a\geq a_f$ and plugs the pair $(a,\lambda_a)$ in (\ref{eff2}). 
The picture, emerging from this analysis, is displayed in Fig. \ref{fig2}. 
There exists a critical value $a_c= -0.4978...$, such that 
$\eta_f(P_{\rm max}) < \eta_{\rm CA}$ for all $a>a_c$. The conventional Schr\"odinger junction $a=0$ is in this range. 
For $a_f\leq a<a_c$ one has  
$\eta_f(P_{\rm max}) > \eta_{\rm CA}$ in some interval of $r$, as shown in Fig. \ref{fig2} for $a=-0.7$. 
Because of (\ref{eff2}) and (\ref{ent2}), $\dot{S}(a_f) = 0$ implies that $\eta_f(P_{\rm max})$ 
equals precisely the Carnot efficiency $\eta_{\rm C}=1-r$ at $a=a_f$. 

{}From Fig. \ref{fig2} one can deduce also that the enhancement 
can be detected in linear response theory (i.e. in the neighborhood of $r=1$) as well. In fact, for $a\not= 0$ 
the associated Onsager matrix is not symmetric, which is a necessary 
condition for enhancement above $\eta_{\rm CA}$. 

{\it Exact efficiency-bosons}: For bosons the computation is totally analogous, except for the presence of a 
singularity in the integrand of (\ref{curr},\ref{en}) at $k^2=-2m(\lambda_i+a)/\beta_i$. In order to exclude it 
from the range of integration, we have to assume $\lambda_i +a >0$. The bosonic 
counterparts of (\ref{pw2},\ref{eff2}) are 
\begin{equation}
\lambda - (1-\e^{\lambda+a}) \ln (1-\e^{-\lambda-a}) =0\, , \quad \lambda +a>0\, ,    
\label{pw2b}
\end{equation} 
\begin{eqnarray}
\eta_b(P_{\rm max}) = \qquad \qquad \qquad \qquad \quad 
\nonumber \\
\frac{(1-r)\lambda_a \ln \left (1-\e^{-\lambda_a-a}\right )}
{(\lambda_a +a+a r)\ln \left (1-\e^{-\lambda_a-a}\right ) -(1+r){\rm Li}_2\left (\e^{-\lambda_a-a}\right )}\, , 
\nonumber \\  
\label{eff2b}
\end{eqnarray}
where $\lambda_a$ satisfies (\ref{pw2b}). 
The study of equation (\ref{pw2b}) shows that for $a<a_b=-0.1792...$ 
there is no (real) solution for $\lambda$. 
There is one solution of (\ref{pw2b}) for $a=a_b$, which is a saddle point of the power $P$. 
In the interval $a_b<a\leq 0$ there are two solutions, one of which being a maximum 
of $P$. Finally, for $a>0$ there is one solution, which also leads to maximal $P$. Summarizing, 
for each $a>a_b$ there exist $\lambda_a$ satisfying both conditions (\ref{pw2b}) and 
giving a maximal power. Moreover, the entropy production (\ref{ent1}) for bosons is positive in this range. 
\begin{figure}[ht]
\begin{center}
\begin{picture}(80,100)(55,11)
\includegraphics[scale=0.68]{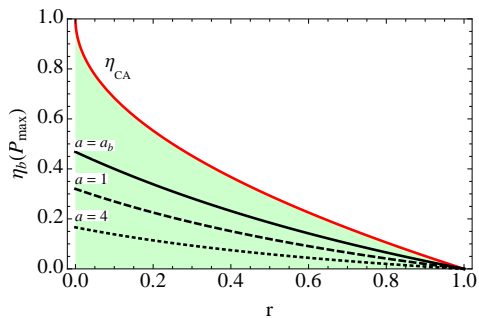}
\end{picture} 
\caption{(Color online) The CA bound (continuous red line) compared to 
$\eta_b(P_{\rm max})$ for $a=4$ (dotted line), $a=1$ (dashed line) and $a=a_b$ (continuous black line).}
\label{fig3}
\end{center}
\end{figure} 

For illustration we have plotted in Fig. \ref{fig3} the efficiency $\eta_b(P_{\rm max})$ 
for some values of the control parameter $a$. It turns out that $\eta_b(P_{\rm max})$ never exceeds 
$\eta_{\rm CA}$ in the allowed domain $a>a_b$. At maximal power 
the bosonic junctions behave therefore differently from the fermionic ones. 
We stress that the condition (\ref{pw2b}) is essential for this conclusion. 
If we release this condition, there exist points in the $(a,\lambda)$-plane (e.g. $(a=-1,\lambda=28)$) with 
positive entropy production, in which also the bosonic efficiency becomes larger then 
$\eta_{\rm CA}$ and approaches $\eta_{\rm C}$. However the power, delivered in these points, is not maximal. 
\begin{figure}[ht]
\begin{center}
\begin{picture}(80,100)(55,11)
\includegraphics[scale=0.68]{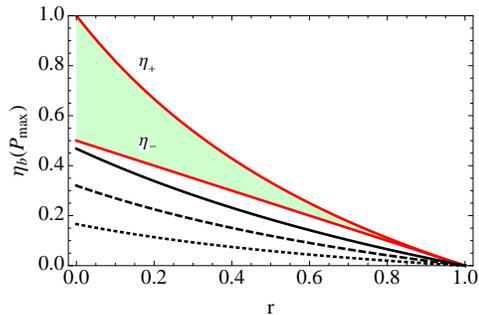}
\end{picture} 
\caption{(Color online) The $\eta_\pm$ bounds (continuous red lines) compared to 
$\eta_b(P_{\rm max})$ for the same values of $a$ as in Fig.\ref{fig3}.}
\label{fig4}
\end{center}
\end{figure} 

{\it Comparison with other bounds}: For classical engines, which can reach the Carnot efficiency $\eta_{\rm C}$ 
in the reversible limit, the following upper and lower bounds 
\begin{equation}
\frac{1}{2}\eta_{\rm C} \equiv \eta_- \leq \eta(P_{\rm max}) \leq \eta_+ \equiv \frac {\eta_{\rm C}}{2-\eta_{\rm C}}\, , 
\label{newbounds}
\end{equation}
have been established in \cite{ss}-\cite{eklb} without referring to linear response theory. For 
comparison with the CA bound we observe that 
$\eta_-\leq \eta_{\rm CA}\leq \eta_+$. Since $\eta_f(P_{\rm max})=\eta_{\rm C}$ for $a=a_f$,  
the fermion efficiency exceeds for appropriate values of $a$ not only $\eta_{\rm CA}$, but also $\eta_+$. For 
bosonic junctions one has instead 
$\eta_b(P_{\rm max})< \eta_-$ for all allowed values $a\geq a_b$, as illustrated in Fig. \ref{fig4}.

{\it Conclusions}: We derived and analyzed systematically the exact efficiency 
$\eta(P_{\rm max})$ for critical Schr\"odinger junctions in the 
Landauer-B\"uttiker steady state. Provided 
that the transmission probability between the two reservoirs does not vanish, 
the intensity of the interaction in the junction 
is irrelevant for $\eta(P_{\rm max})$ in the critical regime. 
Quantum irreversibility is implemented in our framework by a  
spontaneous breaking of time-reversal symmetry. 
We discovered that such a breaking is compatible with vanishing entropy 
production for certain value of the parameter $a$. In fact, in the fermion 
case $\dot{S}(a_f) = 0$, implying that $\eta_f(P_{\rm max})$ reaches 
the Carnot efficiency. The same mechanism works for bosons as well, 
but the corresponding value of $a$ in this case is not in the regime of 
maximal power. Further clarifying the role 
of the parameter $a$ and its impact on other physical observables 
(maximal efficiency, quantum noise,...) 
represents an interesting subject for future investigations.

\end{document}